\documentclass[prd,aps,nofootinbib,tightenlines]{revtex4}

\newcommand{\checked}[1]{}

\usepackage{graphicx}
\usepackage{tabularx}
\usepackage{amsmath}

\newcommand{\beq}{\begin{equation}}
\newcommand{\eeq}{\end{equation}}
\newcommand{\bqa}{\begin{eqnarray}}
\newcommand{\eqa}{\end{eqnarray}}

\begin{document}

\title{Modeling the nonperturbative contributions to the complex heavy-quark potential}
\author{Yun Guo$^{a,b}$, Lihua Dong$^{a,b}$, Jisi Pan$^{a,b}$ and Manoel R. Moldes$^{c}$}
\affiliation{
$^a$Department of Physics, Guangxi Normal University,
Guilin, 541004, China\\
$^b$Guangxi Key Laboratory of Nuclear Physics and Technology, Guilin, 541004, China\\
$^c$Departamento de Fisica de Particulas, Universidade de Santiago de Compostela,
E-15782 Santiago de Compostela, Galicia, Spain}

\begin{abstract}
In this paper, we construct a simple model for the complex heavy quark potential which is defined through the Fourier transform of the static gluon propagator. Besides the hard thermal loop resummed contribution, the gluon propagator also includes a nonperturbative term induced by the dimension two gluon condensate. Within the framework of thermal field theory, the real and imaginary parts of the heavy quark potential are determined in a consistent way without resorting to any extra assumption as long as the exact form of the retarded/advanced gluon propagator is specified. The resulting potential model has the desired asymptotic behaviors and reproduces the data from lattice simulation reasonably well. By presenting a direct comparison with other complex potential models on the market, we find the one proposed in this work shows a significant improvement on the description of the lattice results, especially for the imaginary part of the potential, in a temperature region relevant to quarkonium studies.
\end{abstract}

\maketitle

\section{Introduction}

The heavy-ion experiments at RHIC and the LHC have shown very rich and interesting physics that cannot be interpreted by simple extrapolation from proton-proton collisions, which indicates the formation of a new form of matter --- the quark-gluon plasma (QGP) during the ultrarelativistic heavy-ion collisions. Heavy quarkonium dissociation has been proposed long time ago as a very sensitive probe to study the hot and dense medium\cite{Matsui:1986dk}. Bound states of heavy quarks could survive inside the plasma where the temperature $T$ is higher than the deconfining temperature. However, color screening produced by the light quarks and gluons weakens the interaction between the quark-antiquark pair and leads to the dissociation of quarkonia. Since excited states are more weakly bound than the lower ones, the successive dissociations can possibly serve as a thermometer of QGP\cite{Mocsy:2008eg}.

The studies on quarkonia can be carried out in the nonrelativistic limit due to their large masses, where a quantum mechanical description becomes available. 
As the basic input in the Schr\"odinger equation, the heavy-quark (HQ) potential turns to be very crucial to understand the physical properties of the bound states. At zero temperature, the well-known Cornell potential successfully describes the experimentally observed quarkonium spectroscopy and agrees with the lattice simulations very well. Within the framework of effective field theory (EFT) of QCD, {\em i.e.}, potential nonrelativistic QCD (pNRQCD), substantial development has been achieved in the heavy quarkonium physics\cite{Pineda:1997bj,Brambilla:1999xf}. The EFT was also generalized to finite temperature QCD which justified the description of heavy quarkonia in terms of an in-medium potential. However, the EFT at finite temperature involves much more complications due to the appearance of some extra $T$-dependent scales\cite{Brambilla:2008cx}. As a result, constructing phenomenological potential models has been widely considered over the past decades which provides an alternative way to analyze the in-medium behaviors of the bound states.

In previous studies, the color singlet free energy or internal energy of a static quark pair obtained from lattice simulations was identified with the HQ potential. In addition, based on these lattice results, various proposals for the potential models have also been extensively discussed, see Refs.~\cite{Mocsy:2007jz,Wong:2004zr,Cabrera:2006wh,Alberico:2006vw} for examples. However, the real-valued potential models cannot really represent the HQ potential in the hot medium because it must include an imaginary part induced by the color singlet-octet transition as well as the Landau damping of the low-frequency gauge fields\cite{Brambilla:2011sg}. A first step toward a QCD derivation of the HQ potential at finite temperature was carried out in Ref.~\cite{Laine:2006ns}. In hard thermal loop (HTL) resummed perturbation theory, the static Wilson loops were computed in the imaginary-time formalism. After analytical continuation to Minkowski space, it was found that besides a Debye screened potential as its real part, the potential also contains an imaginary part which determines the decay width of a quarkonium state. Such a perturbative calculation in the weak-coupling limit, however, is only valid when the distance $r$ between the quark and antiquark is small. In the past long period of time, the large distance behavior of the complex potential is not clear due to the lack of the corresponding lattice data. Fortunately, progress has been made in recent years\cite{Rothkopf:2011db,Bazavov:2014kva,Burnier:2014ssa,Burnier:2015tda}. Burnier {\em et al.} have measured the complex-valued static potential by first principle simulations in quenched QCD. In a latest publication\cite{Burnier:2016mxc}, the improved results with reduced finite volume artifacts have been provided.

To sufficiently describe the interaction between the quark pair at finite temperature, there have already been some attempts to develop complex HQ potential models. In Ref.~\cite{Thakur:2013nia}, Thakur {\em et al.} defined the complex HQ potential by Fourier transforming the product of the Cornell potential in momentum space and the inverse dielectric function $\epsilon^{-1}(p)$. Therefore, medium effects are entirely encoded in the complex dielectric function which has been calculated in HTL perturbation theory.  Solving the Schr\"odinger equation with such a complex HQ potential, the binding energies and decay widths of quarkonia have been obtained. However, they did not make a comparison between their potential model and the corresponding lattice results. As we will show later, predictions from this model cannot reproduce the data very well and some asymptotic behaviors are also found to be unphysical. In Ref.~\cite{Burnier:2015nsa}, Burnier {\em et al.}  constructed a complex potential model based on the generalized Gauss law\cite{Dixit:1989vq,Digal:2005ht} and similarly as Ref.~\cite{Thakur:2013nia}, medium effects are incorporated by using the same dielectric function. The predicted imaginary part of the potential based on the model is only satisfactory when $T$ is large and $r$ is small. Therefore, for better understanding the in-medium properties of quarkonia, a more accurate HQ potential model is required which is expected to be in agreement with the lattice data at a quantitative level.

For the above mentioned purpose, the current paper aims to construct a complex HQ potential model which can be used for other phenomenological studies on the heavy quarkonia. The rest of the paper is organized as follows. In Sec.~\ref{pt}, we briefly review the calculation of the complex potential in perturbation theory which provides the Coulombic contribution in our potential model. In Sec.~\ref{geKMS}, we adopt a phenomenological gluon propagator whose nonperturbative term is induced by the dimension two gluon condensate. Performing Fourier transform of such a gluon propagator in Keldysh representation, the obtained HQ potential has a real part which is identical to the Karsch-Mehr-Satz (KMS) potential model. On the other hand, the imaginary part  presents some unexpected features and does not agree with the lattice simulation. Improvements are discussed in Sec.~\ref{imKMS} where, by inspecting the asymptotic behaviors of the model proposed in Sec.~\ref{geKMS}, an additional string contribution is introduced in the gluon propagator. The resulting HQ potential model has been compared to other available models in Refs.~\cite{Thakur:2013nia, Burnier:2015nsa} as well as the lattice results in Ref.~\cite{Burnier:2016mxc}. In a temperature region relevant to quarkonium physics, a significant improvement on the imaginary part of the HQ potential is observed. Finally, we give a short summary in Sec.~\ref{sum}.

\section{Perturbative heavy quark potential at finite temperature}\label{pt}

At zero temperature, the interaction between a static quark pair can be successfully described by the Cornell potential. It takes a form of a Coulomb plus a linear part,
\beq
\label{cor}
V_{\rm Cornell}=-\frac{\alpha_s}{r}+\sigma r\, ,
\eeq
where $\alpha_s=g^2 C_F/(4 \pi)$ is the strong coupling constant, $\sigma$ is the so-called string tension which has the dimension of energy square. At finite temperature, the potential at short distances can be computed in thermal field theory with perturbation expansion. In the real time formalism, the propagator is given by a $2 \times 2$ matrix. It is more convenient to use the Keldysh representation where we have three independent components named retarded ($D_{R}$), advanced ($D_{A}$) and symmetrical ($D_{F}$) propagators. Their relation to the physical ``11" component is given by $D_{11}=(D_{R}+D_{A}+D_{F})/2$. Within hard-thermal-loop approximation, one can compute the self-energy contributions which are used to determine the resummed gluon propagators through the Dyson-Schwinger equation. The perturbative HQ potential $V^{\rm p}$ can be obtained from the following Fourier transform \footnote{From here on, $D$ only denotes the temporal component of the gluon propagator which is relevant to the HQ potential. We introduce a supper script ``p" to indicate perturbative quantities, accordingly a  supper script ``np" stands for the nonperturbative quantities.}
\beq
V^{\rm p}(\hat{r}) = -g^2 C_F \int \frac{d^3 \mathbf{p}}{(2\pi)^3} \left( e^{i {\mathbf{p}} \cdot {\mathbf{r}}}-1 \right){D^{\rm p}_{11}}(p_0=0,{\bf p})\, ,
\eeq
where $D^{\rm p}_{11}$ refers to the physical component of the resummed gluon propagator and $\hat{r}=r m_D$ with the Debye mass given by $m_D^2 = (N_f+2N_c) \frac{g^2 T^2}{6}$. At leading order, the static gluon propagator in the above Fourier transform reads
\bqa
\mathrm{Re}\,{D_{11}^{\rm p}}(p_0=0,{\bf p})&=&D^{\rm p}_{R}(p_0=0,{\bf p})=D^{\rm p}_{A}(p_0=0,{\bf p})=\frac{1}{p^2+m_D^2}\, , \\
\mathrm{Im}\,{D_{11}^{\rm p}}(p_0=0,{\bf p})&=&\frac{1}{2}D^{\rm p}_{F}(p_0=0,{\bf p})= \frac{-\pi T m_D^2}{p(p^2+m_D^2)^2}\, .
\eqa

The real part of the potential is obtained by the Fourier transform of the retarded/advanced propagator while the imaginary part comes from the symmetric propagator in Keldysh representation. Explicitly, we have\cite{Laine:2006ns,Dumitru:2009fy}
\bqa
\label{Vpre}
\mathrm{Re}\,V^{\rm p}(\hat{r})& =& -g^2 C_F \int \frac{d^3 \mathbf{p}}{(2\pi)^3} \left( e^{i {\mathbf{p}} \cdot {\mathbf{r}}}-1 \right) \frac{1}{p^2+m_D^2}=- \alpha_s \bigg(m_D+\frac{e^{-\hat{r}}}{r}\bigg)\, , \\
\label{Vpim}
\mathrm{Im}\,V^{\rm p}(\hat{r})& =& -g^2 C_F \int \frac{d^3 \mathbf{p}}{(2\pi)^3} \left( e^{i {\mathbf{p}} \cdot {\mathbf{r}}}-1 \right) \frac{-\pi T m_D^2}{p(p^2+m_D^2)^2}=- \alpha_s T \phi_2(\hat{r})\, ,
\eqa
where
\begin{equation}
\phi_n(\hat{r}) = 2 \int_0^\infty dz \frac{z}{(z^2 +1)^n} \left[1- \frac{\mathrm{sin}(z \hat{r})}{z \hat{r}}\right]\, .
\end{equation}
Notice that for the real part, the $r$-independent term is divergent and we have subtracted a vacuum contribution $1/p^2$ in the integrand to get a finite result. As compared to the vacuum case, the Coulombic behavior at small distances gets screened and a nonzero imaginary contribution appears. However, the above perturbation theory is not capable of dealing with the medium corrections to the string contribution in the Cornell potential which will be considered by constructing phenomenological models and discussed in the next section.

\section{An extended Karsch-Mehr-Satz heavy-quark potential model}\label{geKMS}

To study the in-medium properties of the heavy bound states, such as charmonia and bottomonia, in the nonrelativistic limit, a proper potential that needs to be specified in the Schr\"odinger equation contains nonperturbative physics due to the typical size of the charm and bottom quark bound states. Therefore, we cannot directly use the above perturbative potential to describe the interactions. In Ref.~\cite{Megias:2005ve}, a new phenomenological term has been added to the perturbative (retarded) gluon propagator $D^{{\rm p}}_R$ in order to account for the effects coming from the low frequency modes incorporated in the dimension two gluon condensates. As a result, the full retarded propagator $D_R$ at static limit takes the following form
\begin{equation}\label{fullDR}
D_R(p_0=0,{\bf p}) \equiv D^{{\rm p}}_R(p_0=0,{\bf p})+D^{\rm np}_R(p_0=0,{\bf p})=
\frac{1}{p^2+m_D^2}+\frac{m_G^2}{(p^2+m_D^2)^2}\, .
\end{equation}
The above equation can be considered as an analogy to the condensates at zero temperature\cite{Chetyrkin:1998yr} which implies a term $m_G^2/p^4$ to be added to the vacuum perturbative gluon propagator $1/p^2$. Here, $m_G^2$ is a dimensional constant. Several applications based on the propagator given in Eq.~(\ref{fullDR}) have been carried out, see Refs.~\cite{Megias:2005ve,Megias:2007pq,Megias:2009mp} for examples. Here, we are interested in the Fourier transform of $D^{\rm np}_R$ at static limit which leads to the following nonperturbative string contribution to the real part of the potential,
\begin{equation}\label{Vnp}
\mathrm{Re}\,V_{\rm I}^{{\rm np}}(\hat{r}) = \frac{\alpha_s m_G^2}{2 m_D}\left[1-\exp\left( -\hat{r} \right)\right]\, .
\end{equation}
We use  $V_{\rm I} = V^{\rm p}+V^{\rm np}_{\rm I}$ to denote the complex HQ potential model discussed in this section. Improvements on $V^{\rm np}_{\rm I}$ will be discussed in Sec.~\ref{imKMS} and the resulting potential model is then denoted as $V_{\rm II} = V^{\rm p}+V^{\rm np}_{\rm II}$. The real part of $V_{\rm I}$ is the sum of Eqs.~(\ref{Vpre}) and (\ref{Vnp}). By matching it onto the Cornell potential at small distances, we find the dimension two constant $m_G^2$ can be related to the string tension through $\sigma=\alpha_s m_G^2/2$. Therefore, $\mathrm{Re}\,V_{\rm I}$ is actually identical to the famous KMS potential model\cite{Karsch:1987pv} in which the large distance interaction is described as a QCD string screened at the same scale as the perturbative contribution. Explicitly, we have
\begin{equation}\label{Vre}
\mathrm{Re}\,V_{\rm I}(\hat{r}) =- \alpha_s \bigg(m_D+\frac{e^{-\hat{r}}}{r}\bigg)+
\frac{\sigma}{m_D}\left[1-\exp\left( -\hat{r} \right)\right]\, .
\end{equation}

Inspired by the above analysis on the real part of the potential, we will also consider adding a string contribution, which describes the large distance behavior of $\mathrm{Im}\,V$, to the perturbative symmetric propagator $D^{\rm p}_{F}$ . In equilibrium, the symmetric propagator can be related to the retarded and advanced ones through the following identity
\begin{equation}\label{Rela}
D_{F}(P)= (1+2 n_B(p_0))\, \mbox{sgn}(p_0)\,[D_{R}(P)-D_{A}(P)]\, ,
\end{equation}
which is valid for full propagators as a consequence of the KMS condition\cite{Carrington:1996rx}. In the above equation, $n_B$ is the Bose-Einstein distribution function and the four-momentum $P\equiv(p_0,{\bf p})$. Although only the static forms of the propagators are required in the Fourier transform, one still need to know the $p_0$-dependent propagators $D_{R}(P)$ and $D_{A}(P)$ in order to compute $D_F(p_0=0,{\bf p})$ through Eq.~(\ref{Rela}). To make it more clear, we consider the distribution function $n_B$ of on-shell thermal gluons in small $p_0$ limit
\begin{equation}\label{dis}
(1+2n_B(p_0))\, \mbox{sgn}(p_0)= \frac{2 T}{p_0}+{\cal O}(p_0^0) \, ,
\end{equation}
which indicates that the leading contribution from $D_{R}(P)-D_{A}(P)$ should be linear in $p_0$ in order to have a nonzero and finite symmetric propagator at static limit. In fact, for the perturbative terms, we have\cite{Dumitru:2009fy}
\begin{equation}\label{retar}
D^{{\rm p}}_{R/A}(P)=(p^2-\Pi_{R/A}(P))^{-1}=\bigg(p^2-m_D^2\bigg(\frac{p_0}{2 p}\ln
\frac{p_0+p\pm i\epsilon} {p_0-p\pm i\epsilon} -1\bigg)\bigg)^{-1}\, ,
\end{equation}
where $\Pi_{R/A}(P)$ is (the temporal component of) the retarded/advanced gluon self-energy at leading order. Performing a Taylor expansion assuming $p_0\rightarrow 0$, it is straightforward to show
\begin{equation}
D^{{\rm p}}_R(P)-D^{{\rm p}}_A(P)=\frac{m_D^2}{2p}\frac{-2\pi
i}{(p^2+m_D^2)^2}\, p_0+{\cal O}(p_0^2)\, .
\end{equation}
Together with Eqs.~(\ref{Rela}) and (\ref{dis}), one can get the symmetric propagator $D^{{\rm p}}_F(p_0=0,{\bf p})$ whose Fourier transform determines $\mathrm{Im}\,V^{\rm p}$ as already calculated in Eq.~(\ref{Vpim}).

However, the $p_0$-dependence of the nonperturbative propagators $D^{{\rm np}}_{R/A}(P)$ is not known. As a ``minimal" extension of the corresponding perturbative result, it is also introduced in a similar way by replacing $m_D^2$ with $-\Pi_{R/A}(P)$ and we assume
\begin{equation}\label{retarnp}
D^{{\rm np}}_{R/A}(P)=m_G^2(p^2-\Pi_{R/A}(P))^{-2}=m_G^2\bigg(p^2-m_D^2\bigg(\frac{p_0}{2 p}\ln
\frac{p_0+p\pm i\epsilon} {p_0-p\pm i\epsilon} -1\bigg)\bigg)^{-2}\, ,
\end{equation}
which has the desired static limit and leads to the following result
\begin{equation}
D^{{\rm np}}_R(P)-D^{{\rm np}}_A(P)=\frac{m_G^2 m_D^2}{p}\frac{-2\pi
i}{(p^2+m_D^2)^3}\, p_0+{\cal O}(p_0^2)\, .
\end{equation}
Accordingly, we can obtain the nonperturbative symmetric propagator through Eq.~(\ref{Rela}) as
\begin{equation}
\label{dfnp}
D^{{\rm np}}_F(p_0=0,{\bf p})=\frac{m_G^2 m_D^2}{p}\frac{-4\pi T
i}{(p^2+m_D^2)^3}\, .
\end{equation}
After Fourier transforming Eq.~(\ref{dfnp}), the string contribution to the imaginary part of the HQ potential is found to be
\begin{equation}
\label{Vnpim}
\mathrm{Im}\,V_{\rm I}^{\rm np}(\hat{r}) = -g^2 C_F \int \frac{d^3 \mathbf{p}}{(2\pi)^3} \left( e^{i {\mathbf{p}} \cdot {\mathbf{r}}}-1 \right) \frac{-2\pi T m_G^2 m_D^2}{p(p^2+m_D^2)^3}=- \frac{4\sigma T}{m_D^2} \phi_3(\hat{r})\,.
\end{equation}
Summing up Eqs.~(\ref{Vpim}) and (\ref{Vnpim}), the full imaginary part in the potential model $V_{\rm I}$ reads
\begin{equation}
\label{Vim}
\mathrm{Im}\,V_{\rm I}(\hat{r})=- \alpha_s T \phi_2(\hat{r}) - \frac{4\sigma T}{m_D^2} \phi_3(\hat{r})\,.
\end{equation}
As an extension of the real-valued KMS model, the complex version $V_{\rm I}=\mathrm{Re}\,V_{\rm I}+i \mathrm{Im}\,V_{\rm I}$ is also called the extended KMS potential model.

It is interesting to see if the above simple model could reproduce the lattice data. To do so, we use the lattice data in quenched QCD from Ref.~\cite{Burnier:2016mxc}. The two parameters $\alpha_s$ and $\sigma$ were assumed to be unchanged in a hot medium, once determined at zero temperature. Due to the absence of a $T=0$ lattice measurement, $\alpha_s=0.272$ and $\sigma=0.215$ ${\rm GeV}^2$ are determined by using the data at $113$ ${\rm MeV}$\cite{Burnier:2016mxc}. Applying these values of $\alpha_s$ and $\sigma$ to the relation $\sigma=\alpha_s m_G^2/2$ leads to a dimension two condensate which coincides with the corresponding lattice simulations\cite{Boucaud:2001st,RuizArriola:2004en,Boucaud:2005rm}. Notice that at short distances, the running of $\alpha_s$ is controlled by the scale $1/r$. Given the shortest quark-pair separation available in the data from Ref.~\cite{Burnier:2016mxc}, which is about $0.1 {\rm fm}$, it turns out that ${\rm Re}\,V$ at short distances can be well described by a naive Cornell potential with fixed coupling constant. On the other hand, with the upcoming high resolution lattice simulations at shorter resolved distances, it is certainly important to take into account the running of $\alpha_s(r)$. Furthermore, although the model study on the Polyakov loop in Ref.~\cite{Megias:2005ve} suggests that the nonperturbative finite temperature condensate is consistent with that at zero temperature, a possible medium dependence of the string tension $\sigma$ cannot be ruled out in principle. However, the exact $T$-dependent form of such a nonperturbative quantity has not been fully clear yet. In this work, we will employ a constant $\sigma$ for simplicity and assume all the medium effects on the HQ potential are encoded in the only free parameter $m_D$ in the model. It is worthwhile to mention that such an assumption can effectively avoid any double counting of the medium effects.

Since the extraction of the imaginary part from lattice simulations gets much more challenging than the real part, the lattice data of $\mathrm{Re}\,V$ is used to determine the Debye mass. In addition, we only consider the lattice data of ${\rm Re}\,V$ up to $1$ ${\rm fm}$ in our fit because in this region of the quark pair separations, the lattice reconstruction is most reliable and the error bars are actually very small. As a crosscheck, the values of $m_D$ from the fit to ${\rm Re}\,V$ will be adopted to evaluate the imaginary part of the potential. The optimized values we obtain for $m_D$ at different temperatures are given in Table~\ref{debyemassfit}.
\setlength{\tabcolsep}{11pt}
\renewcommand{\arraystretch}{1.5}
\begin{table*}[!htbp]
\small
\begin{tabularx}{17cm}{ | c | X | X | X | X | X | X | X | X | X |}
\hline
	$T$ [MeV] & 406 & 369 & 338 & 312 & 290 & 271 & 254 & 226 & 113   \\ \hline
	$V_{\rm I}$: $m_D$ [MeV] & $ 423 $ & $ 258 $ & $ 231 $ & $ 134 $ & $ 87.8 $ & $ 0 $ & $ 0 $ & $ 0 $ & $ 0 $ \\ \hline
\end{tabularx}
\caption{Debye mass extracted from the extended KMS model $V_{\rm I}$ fit to the lattice result for ${\rm Re}\,V$ in Ref.~\cite{Burnier:2016mxc}.}.
\label{debyemassfit}
\end{table*}

A critical behavior is found by inspection of the data and the deconfining temperature $T_c$ is around $290$ ${\rm MeV}$. This is actually consistent with the $T$-dependence of $m_D$ as given in the above table. For temperatures below $T_c$, the values of $m_D$ from the fit turn to be extremely small which are at the order of $10^{-6}$ or even smaller. On the other hand, once the temperature exceeds $T_c$, $m_D$ gets a nonzero value which increases with temperature $T$ as expected. Since we are more interested in the behavior of the HQ potential in the deconfined phase, when $T < T_c$, the values of $m_D$ are simply taken to be zero in Table~\ref{debyemassfit}. As a result, $\hat{r}$ vanishes for finite quark pair separation and $\mathrm{Re}\,V_{\rm I}$ is exactly identical to the vacuum Cornell potential in the confined phase.

\begin{figure}[htbp]
\begin{center}
\includegraphics[width=0.4\textwidth]{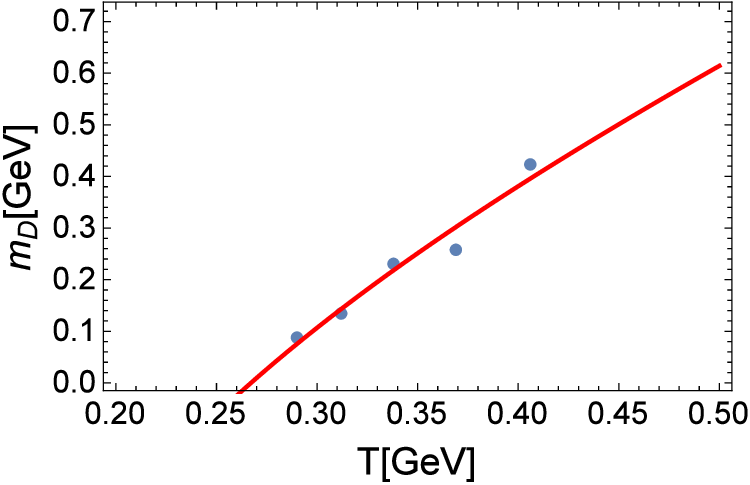}
\caption{Comparison between the parametrization of Debye mass (red solid curve) and its values extracted from the lattice data (blue dots) in Ref.~\cite{Burnier:2016mxc}. }
\label{mdfit}
\end{center}
\end{figure}

As we can see from Table~\ref{debyemassfit}, $m_D$ does not have a simple linear dependence on $T$ which clearly indicates the nonperturbative effects in the temperature region relevant to the quarkonium studies. Interestingly, we find that the extracted Debye mass can be simply parametrized as $m_D(T)=a T+b/T$. Besides the usual leading order result, a new term inversely proportional to $T$ has been included which accounts for the nonperturbative contributions and becomes important when the temperature is decreasing to $T_c$. As shown in Fig.~\ref{mdfit}, in the deconfined phase, the values of $m_D$ can be well reproduced when taking the parameters as $a=1.719$ and $b=-0.123$ ${\rm GeV^2}$. Notice that this simple parametrization of $m_D$ does not apply in the asymptotically high temperature limit because the parameter $a$ is considered as a constant. Furthermore, the negative value of $b$ indicates that the ratio $m_D/T$ decreases as $T$ approaches to $T_c$ from above. The same has also been observed in a massive quasiparticle model when fitting to the equation of state\cite{Peshier:1995ty}. In addition, lattice measurements of the gauge invariant correlation function between Polyakov loops shows that the associated screening mass behaves similarly as that presented in Fig.~\ref{mdfit}\cite{Digal:2003jc,Maezawa:2010vj}. However, based on the two point function of gluons computed on lattice, a contradictory conclusion was obtained where the corresponding gauge dependent mass increases as $T$ approaches to $T_c$\cite{Cucchieri:2001tw,Kaczmarek:2005ui}. Such an upward trend of the ratio $m_D/T$ actually coincides with the result from perturbation calculation \cite{Arnold:1995bh,Ghisoiu:2015uza}. Therefore, as discussed in Ref.~\cite{Dumitru:2010mj}, it would be important to reanalyze the lattice data by using a Higgsed propagator where different modes with
both increasing and decreasing masses are combined.

\begin{figure}[htbp]
\centering
\includegraphics[width=0.27\textwidth]{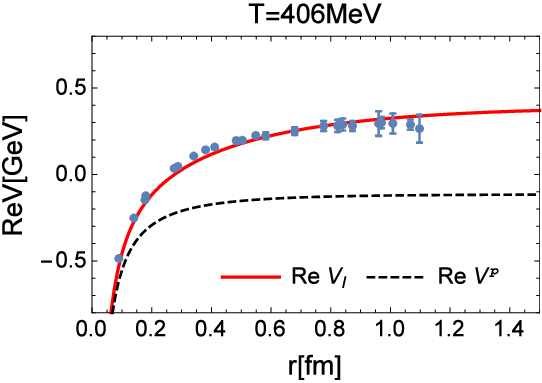}
\includegraphics[width=0.27\textwidth]{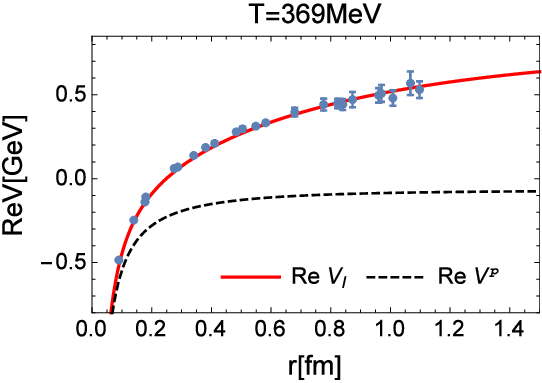}
\includegraphics[width=0.27\textwidth]{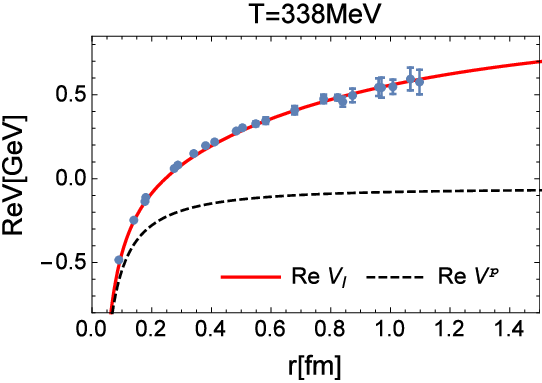}
\includegraphics[width=0.27\textwidth]{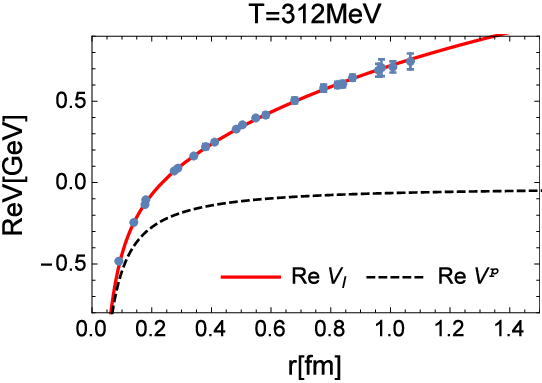}
\includegraphics[width=0.27\textwidth]{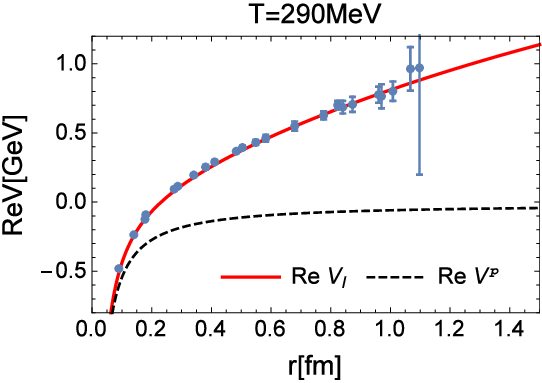}
\includegraphics[width=0.27\textwidth]{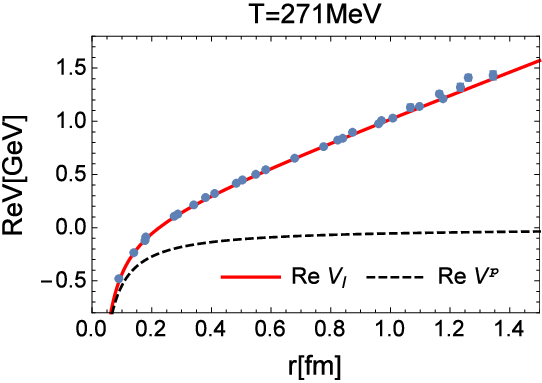}
\vspace*{0.1cm}
\caption{Comparison of $\mathrm{Re}\,V$ between the lattice data in quenched QCD (blue dots) from Ref.~\cite{Burnier:2016mxc} and the extended KMS potential model $V_{\rm I}$ as discussed in Sec.~\ref{geKMS}. The red solid curve denotes the model prediction based on $\mathrm{Re}\,V_{\rm I}$ while the black dashed curve denotes the results from pure perturbative contribution $\mathrm{Re}\,V^{\rm p}$. The critical temperature $T_c=290$ ${\rm MeV}$.}\label{compre1}
\end{figure}

\begin{figure}[htbp]
\centering
\includegraphics[width=0.27\textwidth]{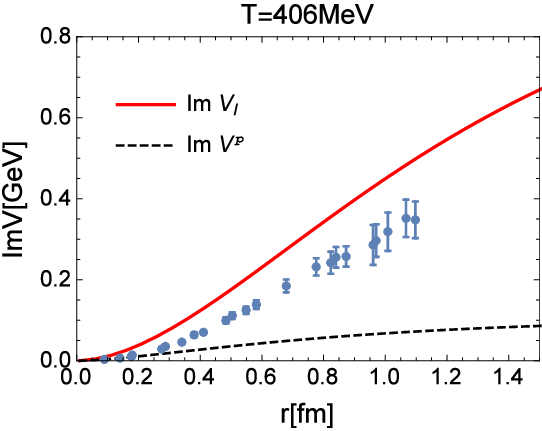}
\includegraphics[width=0.27\textwidth]{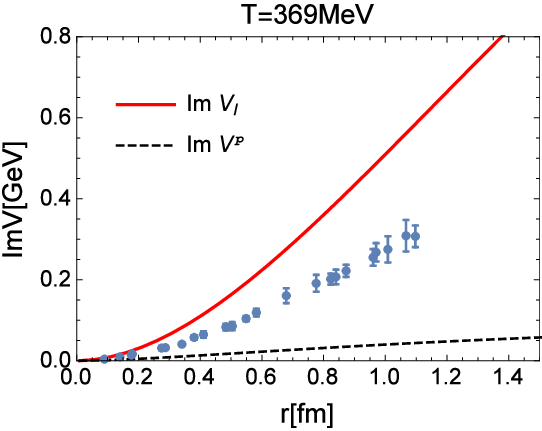}
\includegraphics[width=0.27\textwidth]{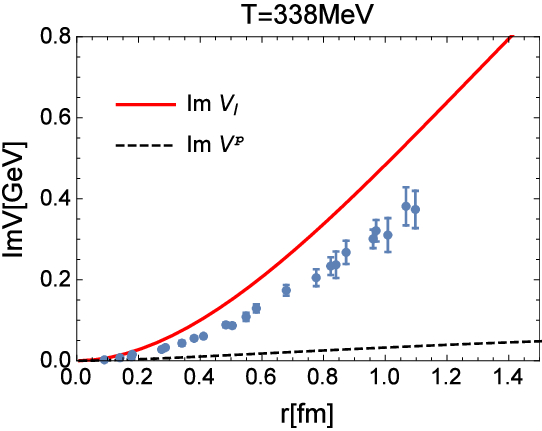}
\includegraphics[width=0.27\textwidth]{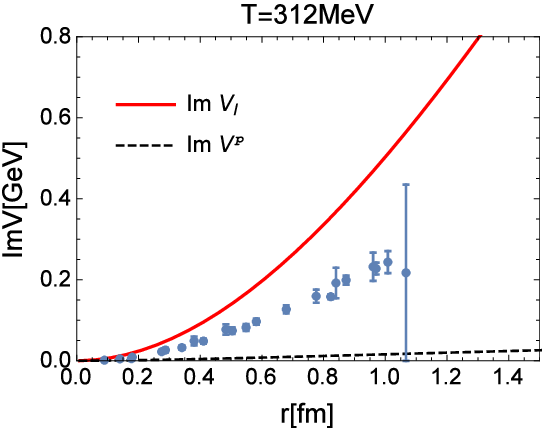}
\includegraphics[width=0.27\textwidth]{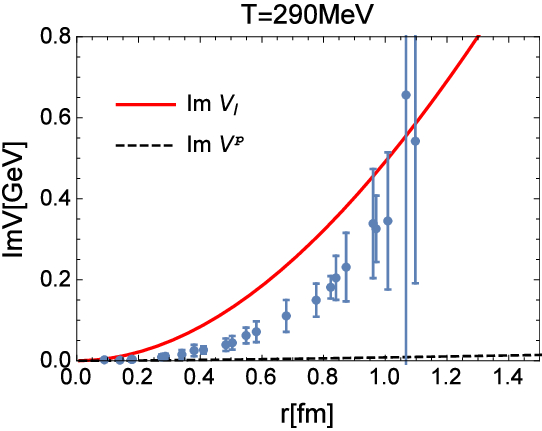}
\includegraphics[width=0.27\textwidth]{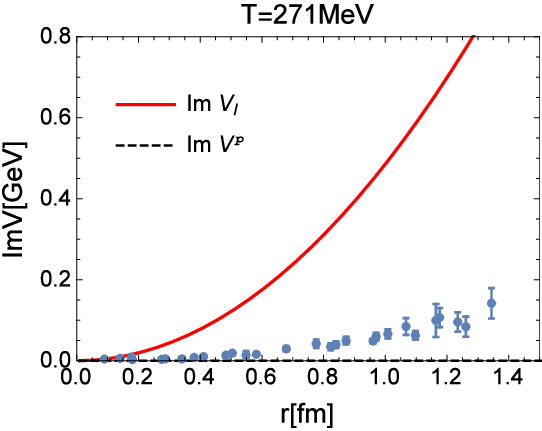}
\vspace*{0.1cm}
\caption{Comparison of $\mathrm{Im}\,V$ between the lattice data in quenched QCD (blue dots) from Ref.~\cite{Burnier:2016mxc} and the extended KMS potential model $V_{\rm I}$ as discussed in Sec.~\ref{geKMS}. The red solid curve denotes the model prediction based on $\mathrm{Im}\,V_{\rm I}$ while the black dashed curve denotes the results from pure perturbative contribution $\mathrm{Im}\,V^{\rm p}$. The critical temperature $T_c=290$ ${\rm MeV}$.}\label{compim1}
\end{figure}

The comparisons between the extended KMS model and the lattice data are given in Fig.~\ref{compre1} for ${\rm Re}\,V$ and in Fig.~\ref{compim1}\footnote{Notice that for the imaginary part of the potential, we actually plot its absolute values in all the figures.} for ${\rm Im}\,V$.  We also plot the pure perturbative results $V^{\rm p}$ which clearly indicate the necessity to include string contributions even for relatively small distances. As shown in Fig.~\ref{compre1}, ${\rm Re}\,V_{\rm I}$ has a good agreement with the lattice data. At very small distances, the Coulombic interaction is dominated while at large distances, it exhibits a screened behavior as suggested by the data. In addition, the lattice data at $T = 271$ ${\rm MeV}$ is nicely reproduced by the vacuum Cornell potential, therefore, our assumption of vanishing Debye mass in the confined phase is justified. On the other hand, ${\rm Im}\,V_{\rm I}$ gets a rapid increase with the quark pair separation which obviously overshoots the lattice data as shown in Fig.~\ref{compim1}. Besides the quantitative deviations in the deconfined phase, a qualitative difference appears when $T < T_c$. Neglecting the cold nuclear effects, in the confined phase, one would expect ${\rm Im}\,V$ is approximately zero which is actually supported by lattice data despite the huge uncertainties. Unfortunately, the model prediction at $T=271$ ${\rm MeV}$ is apparently contradictory to the lattice results. The lack of success of Eq.~(\ref{Vim}) requires improvements on the current potential model, especially for the imaginary part.

Given the model above, it is also important to discuss its asymptotic behaviors which hints at some possible modifications on the extended KMS potential model. In the small distance limit where $\hat{r}\ll 1$, the real part of the potential $\mathrm{Re}\,V_{\rm I}$ reduces to the vacuum Cornell potential and the Coulombic interaction dominates over the string contribution. We can define a distance scale $r_s(T)$ where the nonperturbative effects start to matter. It is determined by requiring $|\mathrm{Re}\,V^{\rm p}(r_s)|=|\mathrm{Re}\,V^{\rm np}(r_s)|$ and we find that $r_s(T)=\sqrt{\alpha_s/\sigma}$. This result is actually $T$-independent because medium effect appears as higher order correction to the Cornell potential when we expand ${\rm Re}\,V_{\rm I}$ with respect to $\hat{r}$. Since this perturbative expansion is valid for $\hat{r}\ll 1$, the above result is applicable when $m_D \ll \sqrt{\sigma/\alpha_s}$ which can be satisfied for not very high temperatures. To study the asymptotic behavior of ${\rm Im}\,V_{\rm I}$ in small $\hat{r}$ limit, we need to expand the following functions\footnote{The expansion of $\phi_4(\hat{r})$ will be used later.},
\bqa
\label{phi2}
\phi_2(\hat{r})& \approx & -\frac{1}{9} \hat{r}^2 (3 \ln \hat{r}-4+3 \gamma_E )\, , \\
\label{phi3}
\phi_3(\hat{r})& \approx & \frac{1}{12} \hat{r}^2 +\frac{1}{900} \hat{r}^4 (15 \ln \hat{r}-23+15 \gamma_E )\, , \\
\label{phi4}
\phi_4(\hat{r})& \approx & \frac{1}{36} \hat{r}^2 -\frac{1}{360} \hat{r}^4\, ,
\eqa
where $\gamma_E$ is the Euler-Gamma constant. The imaginary part of the potential develops a nonzero value at finite temperature and quark pair separation. In general, one can expect that the imaginary part of the potential behaves similarly as the real part, namely, $\mathrm{Im}\,V^{\rm p}$ is dominant at very short distances, when starting to separate the quark pair, the contribution from $\mathrm{Im}\,V^{\rm np}$ gets increased and eventually becomes comparable to $\mathrm{Im}\,V^{\rm p}$ at the same distance scale $r_s(T)\sim \sqrt{\alpha_s/\sigma}$. However, this desired feature does not show up in the analysis based on the above potential model $\mathrm{Im}\,V_{\rm I}$. In fact, the distance scale $r_s(T)$ determined through $|\mathrm{Im}\,V^{\rm p}(r_s)|=|\mathrm{Im}\,V^{\rm np}(r_s)|$ is found to be
\beq
\label{rbar}
r_s(T) \approx \frac{1}{m_D} e^{-\frac{\sigma}{\alpha_s m_D^2}}\, ,
\eeq
which is exponentially suppressed when $m_D \ll \sqrt{\sigma/\alpha_s}$. Therefore, for the imaginary part of the potential, the string contribution becomes important at much smaller distances as compared to the real part. For example, taking $\alpha_s=0.272$ and $\sigma=0.215$ ${\rm GeV}^2$,  we find that $r_s(T)\approx 0.2$ ${\rm fm}$ for $\mathrm{Re}\,V_{\rm I}$ which differs the distance scale for $\mathrm{Im}\,V_{\rm I}$ by orders of magnitude. For some typical value of the Debye mass, $m_D \sim 0.3$ ${\rm GeV}$, $r_s(T)$ is about $10^{-4}$ ${\rm fm}$ for the imaginary part. 

As already mentioned before, $\mathrm{Im}\,V_{\rm I}$ has finite values in the confined phase which increase quickly with the distance $r$. The origin of such an incorrect behavior actually comes from $\mathrm{Im}\,V^{\rm np}_{\rm I}$ in the small $\hat{r}$ limit. One can easily check that for vanishing Debye mass, $|\mathrm{Im}\,V_{\rm I}|$ reduces to $\sigma T r^2/3$ which perfectly reproduces the solid curve in the last plot of Fig.~\ref{compim1}. On the other hand, the above discussed problems can be solved if the leading order contribution in $\mathrm{Im}\,V^{\rm np}$ is proportional to $\hat{r}^4 \ln \hat{r}$ instead of $\hat{r}^2$. As a result, the same distance scale $r_s(T) \sim \sqrt{\alpha_s/\sigma}$ is found for both real and imaginary part of the HQ potential and in the confined phase, $\mathrm{Im}\,V$ also vanishes if $m_D$ is assumed to be zero .

When ${\hat r}\rightarrow \infty$, the asymptotic value of $\mathrm{Re}\,V_{\rm I}$ equals $\sigma/m_D-\alpha_s m_D$. 
In general, the Debye screening mass increases with $T$, therefore, $\mathrm{Re}\,V_{\rm I}(\hat{r}\rightarrow \infty)$ decreases as $T$ is getting larger. This is qualitatively in agreement with that suggested by lattice data. In addition, for the imaginary part, we have
\beq
\label{Viminfty}
\mathrm{Im}\,V_{\rm I}(\hat{r}\rightarrow \infty)= - \alpha_s T - \frac{2\sigma T}{m_D^2}\, .
\eeq
According to this equation, the asymptotic value of $\mathrm{Im}\,V_{\rm I}$ could have a nontrivial dependence on the temperature $T$. Only at very high temperatures where the Debye mass $m_D \sim T$, we can expect $|\mathrm{Im}\,V_{\rm I}(\hat{r}\rightarrow \infty)|$ increase with increasing $T$ provided $m_D > \sqrt{2\sigma/\alpha_s}$. However, the current lattice simulations on the complex HQ potential cannot provide us sufficient information about the asymptotic values of $\mathrm{Im}\,V$ at large $\hat{r}$.

\section{An improved Karsch-Mehr-Satz heavy-quark potential model}\label{imKMS}

For the purpose of quantitatively describing the lattice data, in this section, we will discuss the improvements on the extended KMS potential model $V_{\rm I}$ as proposed in Sec.~\ref{geKMS}. In fact, the analysis on the asymptotic behavior of $\mathrm{Im}\,V_{\rm I}$ suggests the leading order contribution from $\mathrm{Im}\,V^{\rm np}$ should behave like $\sim \hat{r}^4 \ln \hat{r}$ when $\hat{r}\rightarrow 0$. Therefore, an extra nonperturbative term could be introduced in the symmetric propagator and the resulting contribution to $\mathrm{Im}\,V$ is expected to cancel the $\sim \hat{r}^2$ term in $\mathrm{Im}\,V_{\rm I}^{\rm np}$ in the small $\hat{r}$ limit. This can be achieved in a consistent way through Eq.~(\ref{Rela}) and the key point is to find a proper string contribution which needs to be added to the retarded/advanced propagator $D^{\rm np}_{R/A}$ in Eq.~(\ref{fullDR}).

At finite temperature, a nonperturbative term $m_G^2/(p^2+m_D^2)^2$ in the retarded propagator was introduced based on the extension of the vacuum dimension two gluon condensates. From a  phenomenological point of view, however, we cannot rule out some other possible forms, for example, adding a term $\sim m_G^2 m_D^2/(p^2+m_D^2)^3$ in Eq.~(\ref{fullDR}) does not ruin the vacuum limit $m_G^2/p^4$ since this term vanishes as $m_D\rightarrow 0$. Furthermore, both terms produce the same kind of condensate related to $m_G^2$ and in the Gaussian-like approximation,  lead to the same nonperturbative contribution $\sim 1/T^2$ to the (logarithm of the) Polyakov loop in deconfined phase\cite{Megias:2007pq}. Therefore, we formally write the improved retarded propagator as $\tilde{D}_R \equiv D_R+\delta D_R$ and (the static limit of) the newly added string contribution $\delta D_R$ can be written as
\beq
\label{modDR2}
\delta D_R(p_0=0,{\bf p})\equiv a \frac{m_G^2 m_D^2}{(p^2+m_D^2)^3}\, ,
\eeq
where $a$ is a dimensionless constant. Fourier transforming Eq.~(\ref{modDR2}), an extra contribution to the real part of the potential reads
\beq
\label{coVre}
\mathrm{Re}\,\delta V ({\hat r})=\frac{a}{4}\frac{\sigma}{m_D}(1-e^{-\hat{r}}-\hat{r}e^{-\hat{r}})\, .
\eeq
Qualitatively, the asymptotic behavior of $\mathrm{Re}\,V$ is not affected by the above contribution. Since Eq.~(\ref{coVre}) vanishes when taking $\hat{r}\rightarrow 0$, one still gets the Cornell potential in this limit. On the other hand, the asymptotic value at infinitely large $\hat{r}$ is changed into $(1+a/4)\sigma/m_D-\alpha_s m_D$.

In vacuum, the main contribution to $\mathrm{Re}\,V$ is dominated by $\sigma r$ for large quark pair separation. For $0<T<T_c$, $\mathrm{Re}\,V$ is very close to the Cornell potential as observed by the lattice simulation. Therefore, the medium effects in the confined phase, which strictly speaking are not exactly zero, can be treated perturbatively by assuming ${\hat r}\ll1$. 
For large but finite quark pair separation $r$, we can always assume ${\hat r}\ll 1$ in the confined phase due to $m_D\rightarrow 0$. Since the potential $\mathrm{Re}\,V$ at finite $T$ cannot overshoot the vacuum potential, such medium effects should set in as a negative correction to $\sigma r$. According to Eq.~(\ref{Vnp}), the leading order correction equals $-\sigma \hat{r}^2/(2m_D)$ which is negative as expected. After including the extra contribution in Eq.~(\ref{coVre}), this correction becomes $-\sigma \hat{r}^2/(2m_D)+a \sigma \hat{r}^2/(8m_D)$. To keep it nonpositive, we choose the maximum value of $a$, {\em i.e.}, $a=4$ and it is expected to give the ``most confining" potential. Accordingly, the leading order correction appears at higher order in $\hat{r}$ which is still negative and given by $-\sigma \hat{r}^3/(6 m_D)$. In the above discussion, we ignore the medium correction coming from the perturbative terms because it is small as compared to the corresponding nonperturbative correction when $r$ is large. With the above choice of the constant $a$, the asymptotic value $\mathrm{Re}\,V^{\rm np}({\hat r}\rightarrow \infty)$ is changed from $\sigma/m_D$ to $2\sigma/m_D$ which is identical to other potential models discussed in Refs.~\cite{Dumitru:2009ni,Thakur:2013nia}.

With the improved propagator $\tilde{D}_{R/A}$, we will also study the corresponding changes in the imaginary part of the potential. To do so,  the $p_0$-dependence should be properly introduced in $\delta D_{R/A}$. Here, we adopt the following assumption for the $p_0$-dependent $\delta D_{R/A}$
\beq
\label{retarex}
\delta D_{R/A}(P)=b \frac{ m_G^2 m_D^2}{(p^2-\Pi_{R/A}(P))^{3}}+b^\prime \frac{ m_G^2 (-m_D^2-\Pi_{R/A}(P))}{(p^2-\Pi_{R/A}(P))^{3}}\, .
\eeq
There is a subtlety in the above equation due to the fact that the term with the dimensionless constant $b^\prime$ vanishes in the static limit. Therefore, the recovery to Eq.~(\ref{modDR2}) when $p_0=0$ which requires the dimensionless constant $b$ to be equal to $a$, however, does not impose any constraint on the value of $b^\prime$. In fact, Eq.~(\ref{retarex}) can be considered as a generalized expression of $\delta D_{R/A}(P)$ as compared to its simplest form one could imagine $\sim m_G^2\Pi_{R/A}(P)/(p^2-\Pi_{R/A}(P))^3$. The latter is obtained from Eq.~(\ref{modDR2}) by replacing $m_D$ with the gluon self-energy $-\Pi_{R/A}(P)$ and identical to our assumption Eq.~(\ref{retarex}) only when $b=b^\prime$. The necessity of considering a more general form of $\delta D_{R/A}$, as we will see later, is based on the fact that $b^\prime$ has to take some value different from $b$ in order to meet the crucial requirement on $\mathrm{Im}\,V^{\rm np}$, namely, its leading order contribution should be proportional to $\sim \hat{r}^4 \ln \hat{r}$ in small $\hat{r}$ limit.

Using Eq.~(\ref{Rela}), we can calculate the extra contribution to the symmetric propagator induced by Eq.~(\ref{retarex}),
\begin{equation}
\delta D_F(p_0=0,{\bf p})=-2\pi T i\frac{m_G^2 m_D^2}{p}\bigg[b \frac{3m_D^2}{(p^2+m_D^2)^4}-b^\prime\frac{1}{(p^2+m_D^2)^3}\bigg]\, ,
\end{equation}
which after performing Fourier transform, gives rise to the following correction to the imaginary part of the HQ potential $\mathrm{Im}\,V_{\rm I}^{\rm np}$,
\bqa
\label{coVnp}
\mathrm{Im}\,\delta V(\hat{r}) &=& -b\frac{6 \sigma T }{m_D^2} \phi_4(\hat{r})+ b^\prime \frac{2 \sigma T }{m_D^2}  \phi_3(\hat{r}) \, , \\ \nonumber
&\approx& - b \frac{\sigma T }{m_D^2} \bigg(\frac{{\hat r}^2}{6}-\frac{{\hat r}^4}{60} \bigg)+b^\prime \frac{\sigma T  }{m_D^2} \bigg(\frac{{\hat r}^2}{6}-\frac{23-15\gamma_E-15 \ln {\hat r}}{450} {\hat r}^4\bigg)\, ,\quad\quad\quad {\rm for} \quad {\hat r}\ll 1\,.
\eqa
Here, the small ${\hat r}$ expansion is obtained by using Eqs.~(\ref{phi3}) and (\ref{phi4}). As we can see the leading order contribution from the $\phi_3(\hat{r})$ term is $\sim \hat{r}^2$ and the same holds for the $\phi_4(\hat{r})$ term but with opposite sign. If the values of $b$ and $b^\prime$ were chosen to be the same, the leading order contribution from Eq.~(\ref{coVnp}) would be proportional to $\sim \hat{r}^4 \ln \hat{r}$ and there is no way to cancel the $\sim \hat{r}^2$ term in $\mathrm{Im}\,V_{\rm I}^{\rm np}$. On the other hand, the desired result can be obtained when the relation between the two dimensionless constants $b^\prime-b=2$ is satisfied. After including the correction in Eq.~(\ref{coVnp}), the nonperturbative contribution to $\mathrm{Im}\,V$ now takes the form
\bqa
\mathrm{Im}\,V^{\rm np}_{\rm II}(\hat{r})&=&\mathrm{Im}\,V^{\rm np}_{\rm I}(\hat{r})+\mathrm{Im}\,\delta V(\hat{r})= b\frac{2 \sigma T}{m_D^2} [\phi_3(\hat{r})-3 \phi_4(\hat{r})]\, , \\ \nonumber
&\approx& b\frac{\sigma T }{m_D^2} \frac{30 \ln {\hat r}-31+30\gamma_E}{900} {\hat r}^4\, ,\quad\quad\quad {\rm for} \quad {\hat r}\ll 1\,.
\eqa
In the above equation, $b^\prime$ has already been replaced by $b+2$. Furthermore, in order to have the ``most confining" potential, the constant $b$ is uniquely determined as $b=a=4$.

The above discussion clearly demonstrates the rationality of using the more general assumption Eq.~(\ref{retarex}) for the $p_0$-dependence of $\delta D_{R/A}(P)$. However, one may wonder that what would happen if we added more than one possible term like $\sim m_G^2(\Pi_{R/A})^n/(p^2-\Pi_{R/A})^{n+2}$ (with $n=1,2,\cdots $) to $D^{{\rm np}}_{R/A}(P)$. By inspecting the asymptotic behavior of the real part of the potential, no qualitative change was found when two or more terms are added simultaneously. However, in small $\hat{r}$ limit, the correction to the imaginary part induced by each individual term is subleading with respect to the term $\sim {\hat r}^2$ in $\mathrm{Im}\,V_{\rm I}$, as a result, no cancellation could happen. Unavoidably, we need to use the same trick as Eq.~(\ref{retarex}) to split at least one added term into two parts with different coefficients. Therefore, adding more terms turns to be not helpful which on the other hand, makes the model complicated.

Now, we are ready to write down the improved KMS potential model $V_{\rm II}$ which, as compared to the extended KMS model $V_{\rm I}$ in Sec.~\ref{geKMS}, contains the corrections from Eqs.~(\ref{coVre}) and (\ref{coVnp}). Explicitly, the results are listed below
\bqa
\label{imV}
\mathrm{Re}\,V_{\rm II}(\hat{r}) &=&- \alpha_s \bigg(m_D+\frac{e^{-\hat{r}}}{r}\bigg)+
\frac{2 \sigma}{m_D}\left[1-\exp\left( -\hat{r} \right)\right]-\frac{\sigma}{m_D}\hat{r} \exp\left( -\hat{r} \right)\, , \\ \nonumber
\mathrm{Im}\,V_{\rm II}(\hat{r}) &=&- \alpha_s T \phi_2(\hat{r}) +\frac{8\sigma T}{m_D^2} \phi_3(\hat{r})- \frac{24\sigma T}{m_D^2} \phi_4(\hat{r})\,.
\eqa
The asymptotic values of the above potential model at $\hat{r}\rightarrow 0$ are found to be
\beq
\label{newasy1}
\mathrm{Re}\,V_{\rm II}(\hat{r}\rightarrow 0)= - \alpha_s/r+\sigma r \,  \quad\quad{\rm and}\quad\quad \mathrm{Im}\,V_{\rm II}(\hat{r}\rightarrow 0)= \frac{\alpha_s T}{3} \hat{r}^2 \ln \hat{r} +  \frac{2 \sigma T}{15 m_D^2} \hat{r}^4 \ln \hat{r}\, .
\eeq
Correspondingly, for $\hat{r}\rightarrow \infty$ we have
\beq
\mathrm{Re}\,V_{\rm II}(\hat{r}\rightarrow \infty)= - \alpha_s m_D+\frac{2 \sigma}{m_D} \,  \quad\quad{\rm and}\quad\quad \mathrm{Im}\,V_{\rm II}(\hat{r}\rightarrow \infty)= - \alpha_s T - \frac{4 \sigma T}{m_D^2}\, .
\eeq
According to Eq.~(\ref{newasy1}), now the distance scale $r_s(T)$ is at the order of $\sim \sqrt{\alpha_s/\sigma}$ for both real and imaginary part of the HQ potential. Furthermore, at infinitely large $\hat{r}$, the nonperturbative contribution equals $(1+a/4)\sigma/m_D$ for the real part and $ - b \sigma T/m_D^2$ for the imaginary part. Therefore, $V_{\rm II}^{\rm np}(\hat{r}\rightarrow \infty)$ is determined solely by the dimensionless constant $a$ ($a$=$b$ is required).

Before we show the comparison between the improved KMS potential model $V_{\rm II}=\mathrm{Re}\,V_{\rm II}+i \mathrm{Im}\,V_{\rm II}$ and lattice data,  it is also worthwhile to mention other phenomenological models which have been studied in Refs.~\cite{Thakur:2013nia,Burnier:2015nsa}. We refer to the one in Ref.~\cite{Thakur:2013nia} as Thakur-Kakade-Patra(TKP) model. Accordingly, the model in Ref.~\cite{Burnier:2015nsa} is referred to as Burnier-Rothkopf(BR) model. Explicit forms of these two potential models can be found in the above mentioned references. Although the basic ideas of model construction are very different from each other, the perturbative terms in these models are all expressed by the leading order HTL result, {\em i.e.}, Eqs.~(\ref{Vpre}) and (\ref{Vpim}). On the other hand, despite owning different nonperturbative forms, their asymptotic behaviors of $\mathrm{Re}\,V^{\rm np}$ are actually very similar. Taking $\hat{r}\rightarrow 0$, $\mathrm{Re}\,V^{\rm np}$ reduces to the linear rising Cornell potential. At infinitely large $r$, the asymptotic value $\mathrm{Re}\,V^{\rm np}(\hat{r}\rightarrow \infty)$ obtained from TKP model coincides with the improved KMS model. While for the BR model, the corresponding value becomes $\sim \sigma^{3/4}/\sqrt{m_D}$ which may indicate a different $T$-dependence as compared to the other two models\footnote{At relatively large distances ($\hat{r} \gg 1$), the real parts of the improved KMS model and the BR model decay exponentially, however, the TKP model retains an $\sim 1/r$ behavior.}.

We would like to also mention that there exist some differences among these potential models when we consider the medium effect as a perturbation to the Cornell potential. In the deconfined phase where $m_D$ is very small, the potential can be expanded in term of $\hat{r}$. As just discussed before, in the improved KMS model, the medium correction from $\mathrm{Re}\,V^{\rm np}$ is $-\sigma {\hat r}^3/(6m_D)$, comparing with that from $\mathrm{Re}\,V^{\rm p}$ which is $-\alpha_s m_D {\hat r}/2$, we find a critical distance $\sim \sqrt{\alpha_s/\sigma}$ above which the leading order medium correction is from $\mathrm{Re}\,V^{\rm np}$, while in the region where $r$ is smaller than the critical distance, the correction from $\mathrm{Re}\,V^{\rm p}$ is dominated. On the contrary, the nonperturbative correction due to medium effect is $\sim {\hat r}^2$ in TKP model and $\sim (r \mu)^4$ in BR model with $\mu \sim (m_D^2 \sigma/\alpha_s)^{1/4}$. As a result, for $m_D\rightarrow 0$, the leading order correction comes from the nonperturbative terms even at very small distances. Despite such a qualitative difference, all these medium corrections have negative contributions and none of them overshoots the vacuum potential.

As for the nonperturbative terms in $\mathrm{Im}\,V$, the improved KMS model and BR model share some common features, namely, the Coulombic HTL part dominates at small distances while at asymptotically large distances, the string contribution saturates to some constant as required. As pointed in Ref.~\cite{Burnier:2015nsa}, when $r$ is small, the string term in the BR model rises according to $r^3$ which is subleading with respect to the Coulombic contribution and the corresponding distance scale $r_s(T)$ is comparable to that in $\mathrm{Im}\,V_{\rm II}$. The TKP model, on the other hand, shows some unexpected differences. In small $\hat{r}$ limit, the leading order contribution from the nonperturbative term is proportional to $\hat{r}^2$, so the string part would contribute equally as the Coulombic term even at very small $r$. This is exactly the same as the previous discussed potential model in Eq.~(\ref{Vnpim}).  As a result, $\mathrm{Im}\,V$ in this model gets an unwanted increase proportional to $r^2$ when the temperature is below $T_c$. 
Finally, we find that $\mathrm{Im}\,V^{\rm np}(\hat{r}\rightarrow \infty)$ in TKP model does not converge to some constant, instead a logarithmic divergence $\sim \ln \hat {r}$ exists.

Similar as what we did in Sec.~\ref{geKMS}, the strong coupling constant $\alpha_s$ and the string tension $\sigma$ are assumed to be $T$-independent and the lattice data of the real part of the potential is used to extract the only free parameter $m_D$ in the above models. The corresponding results can be found in Table~\ref{debyemassfit2}. We point out that the value of $m_D$ at a given temperature varies according to the different forms of $\mathrm{Re}\,V$ under consideration, however, the $T$-dependence of $m_D$ in these models looks very similar and all of them can be well described by using the previous parametrization $a T+b/T$. For the improved KMS model, the set of parameters are found to be $a=2.69,b=-0.146$ ${\rm GeV}^2$. We get $a=2.32,b=-0.161$ ${\rm GeV}^2$ and $a=2.38,b=-0.152$ ${\rm GeV}^2$ for the TKP model and BR model, respectively. In all the cases, a negative value of $b$ suggests the downward trend of the ratio $m_D/T$ as $T$ approaches to $T_c$ should be a universal behavior. Here, we will not show the comparisons between the extracted values of $m_D$ and their parametrizations since the outcomes are almost the same as Fig.~\ref{mdfit}.
\setlength{\tabcolsep}{11pt}
\renewcommand{\arraystretch}{1.5}
\begin{table*}[!htbp]
\small
\begin{tabularx}{17cm}{ | c | X | X | X | X | X | X | X | X | X |}
\hline
	$T$ [MeV] & 406 & 369 & 338 & 312 & 290 & 271 & 254 & 226 & 113   \\ \hline
	$V_{\rm II}$: $m_D$  [MeV] & $ 754 $ & $ 546 $ & $ 508 $ & $ 361 $ & $ 278 $ & $ 0 $ & $ 0 $ & $ 0 $ & $ 0 $ \\ \hline
	${\rm TKP\, Model}$: $m_D$  [MeV] & $ 576 $ & $ 365 $ & $ 329 $ & $ 196 $ & $ 129 $ & $ 0 $ & $ 0 $ & $ 0 $ & $ 0 $ \\ \hline
	${\rm BR\, Model}$: $m_D$  [MeV] & $ 603 $ & $ 430 $ & $ 367 $ & $ 273 $ & $ 150 $ & $ 14 $ & $ 0 $ & $ 0 $ & $ 0 $ \\ \hline	
\end{tabularx}\caption{Debye mass extracted from the potential models fit to the lattice result for ${\rm Re}\,V$ in Ref.~\cite{Burnier:2016mxc}. Besides the improved KMS model $V_{\rm II}$, we also extract the Debye mass from TKP model fit to the data. For completeness, the values of Debye mass in BR model obtained in Ref.~\cite{Burnier:2016mxc} are also listed in the table.}.
\label{debyemassfit2}
\end{table*}

\begin{figure}[htbp]
\centering
\includegraphics[width=0.27\textwidth]{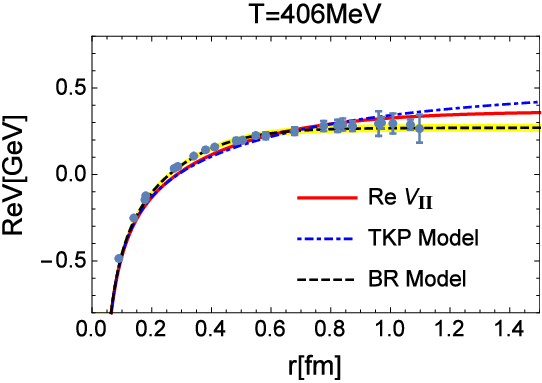}
\includegraphics[width=0.27\textwidth]{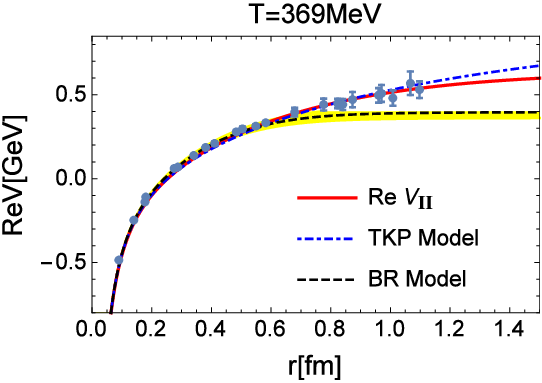}
\includegraphics[width=0.27\textwidth]{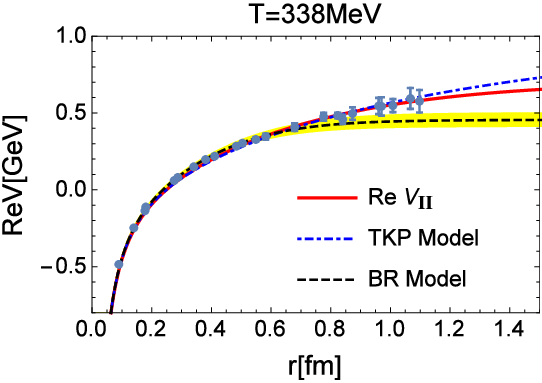}
\includegraphics[width=0.27\textwidth]{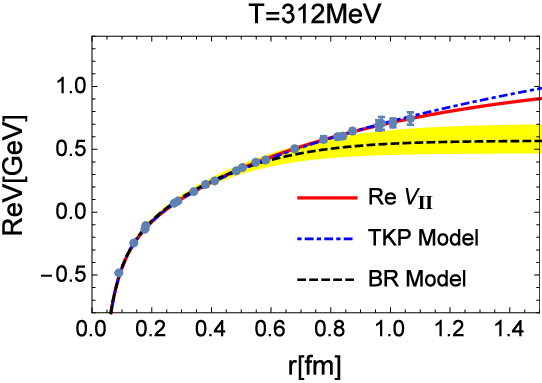}
\includegraphics[width=0.27\textwidth]{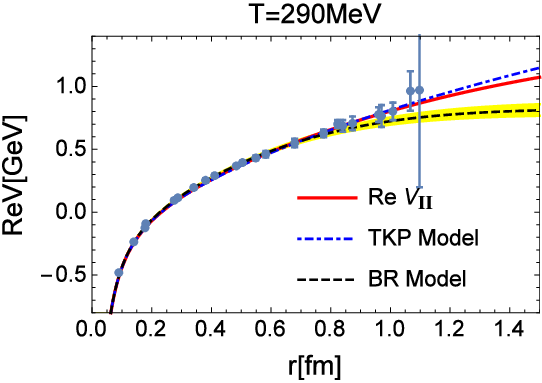}
\includegraphics[width=0.27\textwidth]{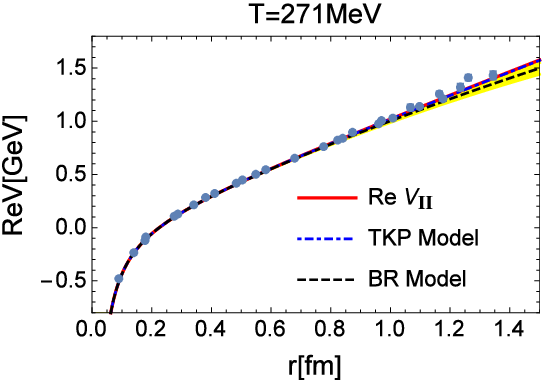}
\vspace*{0.1cm}
\caption{Comparison of $\mathrm{Re}\,V$ between the lattice data in quenched QCD (blue dots) from Ref.~\cite{Burnier:2016mxc} and various complex potential models. The results from the improved KMS potential model $V_{\rm II}$ is denoted by the red solid curve and the blue dash-dotted curve denotes the TKP model. The prediction from BR model is also shown in this figure which is denoted by the black dashed curve with the colored error-bands from the uncertainty in the determination of $m_D$. The critical temperature $T_c=290$ ${\rm MeV}$.}\label{compre2}
\end{figure}

In Fig.~\ref{compre2}, we show the comparison between the model predictions and the lattice data for $\mathrm{Re}\,V$. Besides the improved KMS model $\mathrm{Re}\,V_{\rm II}$, the results obtained from the TKP model and BR model are also plotted in this figure. In fact, the prediction from $\mathrm{Re}\,V_{\rm II}$ is quantitatively the same as compared to the TKP model up to the distances around $1$ ${\rm fm}$. In the following, we only concentrate on the comparison between $\mathrm{Re}\,V_{\rm II}$ and the BR model. Roughly speaking, both models can well reproduce the data. In the confined phase, since the Debye mass is approximately zero, we actually have the Cornell potential and nothing changes as compared to the extended KMS model as discussed in Sec.~\ref{geKMS}. Above the critical temperature, both models behave qualitatively the same, namely, a Debye screened contribution at small distances and a screened string contribution at large distances. In addition, for temperatures slightly above $T_c$, a better agreement can be obtained by using the improved KMS model since it exhibits an upward trend at large distances which is in accordance with the data. On the other hand, at relatively high temperatures, small deviations from data at intermediate and large distances appear in $\mathrm{Re}\,V_{\rm II}$ while the BR model turns to work very well.

\begin{figure}[htbp]
\begin{center}
\includegraphics[width=0.4\textwidth]{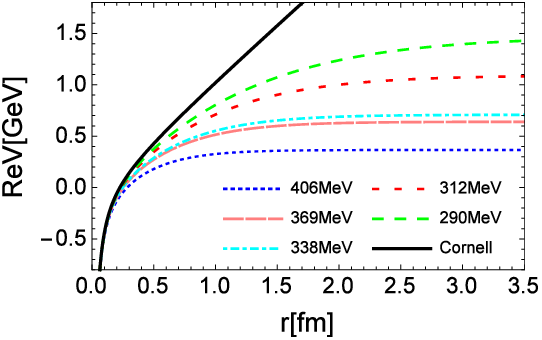}
\caption{The real part of the potential predicted by the improved KMS model $\mathrm{Re}\,V_{\rm II}$ at different temperatures in the deconfined phase. The solid curve denotes to the vacuum Cornell potential. }
\label{larger}
\end{center}
\end{figure}

We consider a temperature regime in the deconfined phase from $T_c=290$ ${\rm MeV}$ to $1.4T_c = 406$ ${\rm MeV}$. As we can see from Fig.~\ref{compre2}, the real part of the potential saturates at $r\approx 1$ ${\rm fm}$ when $T= 406$ ${\rm MeV}$ which demonstrates the screening behavior. On the other hand, for temperatures slightly higher than $T_c$, the potential curve starts to flatten at even larger distances. The distance where $\mathrm{Re}\,V$ saturates strongly depends on the temperature. This qualitatively agrees with the lattice simulations and can be verified by Fig.~\ref{larger} which clearly shows the desired screening behavior of the improved KMS model at very large distances. For example, when $T= 338$  ${\rm MeV}$, $\mathrm{Re}\,V_{\rm II}$ arrives at the asymptotic value at $r\approx 1.5$ ${\rm fm}$.

\begin{figure}[htbp]
\begin{center}
\includegraphics[width=0.4\textwidth]{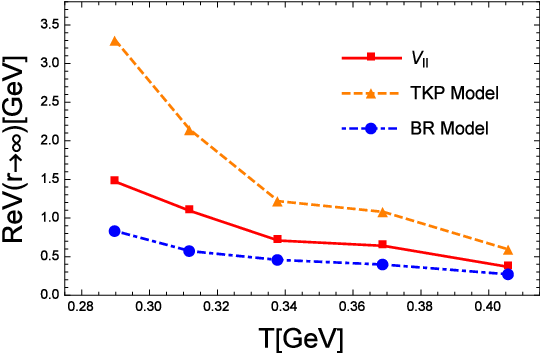}
\caption{Comparison of the asymptotic values $\mathrm{Re}\,V(r \rightarrow \infty)$ from different potential models. }
\label{Vinf}
\end{center}
\end{figure}

According to Fig.~\ref{compre2}, the real part of the potential seems not very sensitive to the exact forms we used in the fit because all the potential models seem to work reasonably well. On the other hand, the discrepancy of the Debye mass among different models leads to very different asymptotic values $\mathrm{Re}\,V(r \rightarrow \infty)$ as shown in Fig.~\ref{Vinf}. Therefore, the binding energies of quarkonia evaluated based on these potential models may differ dramatically. The elimination of such an ambiguity on $\mathrm{Re}\,V(r \rightarrow \infty)$ requires lattice simulations at even larger distances.

\begin{figure}[htbp]
\centering
\includegraphics[width=0.27\textwidth]{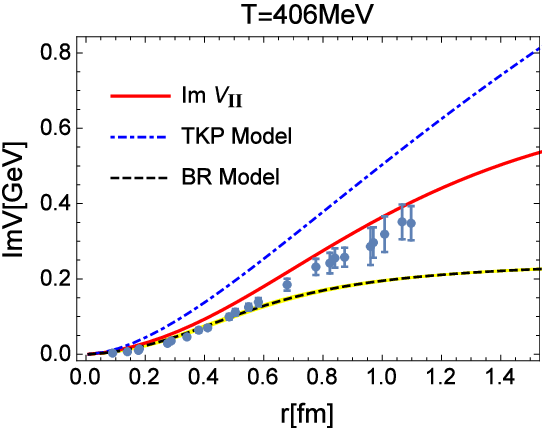}
\includegraphics[width=0.27\textwidth]{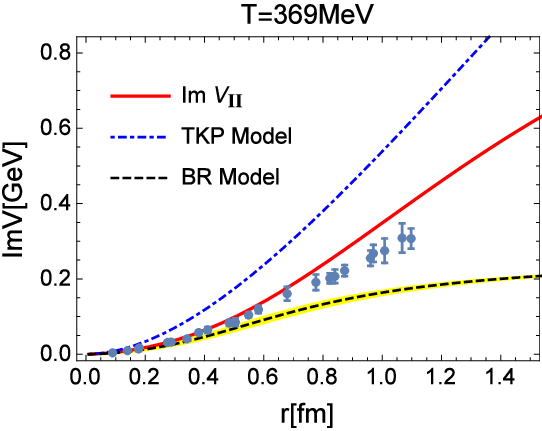}
\includegraphics[width=0.27\textwidth]{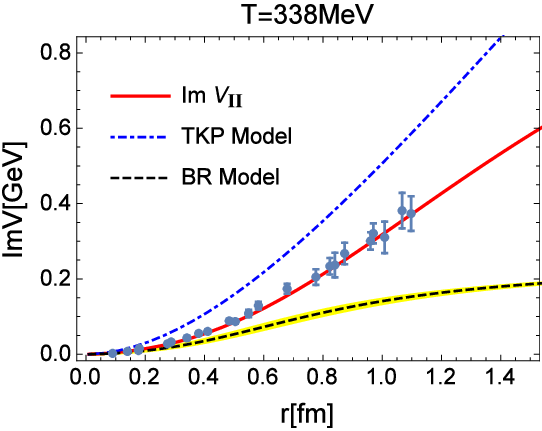}
\includegraphics[width=0.27\textwidth]{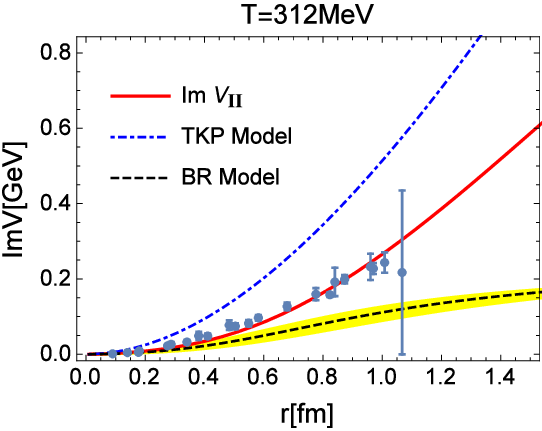}
\includegraphics[width=0.27\textwidth]{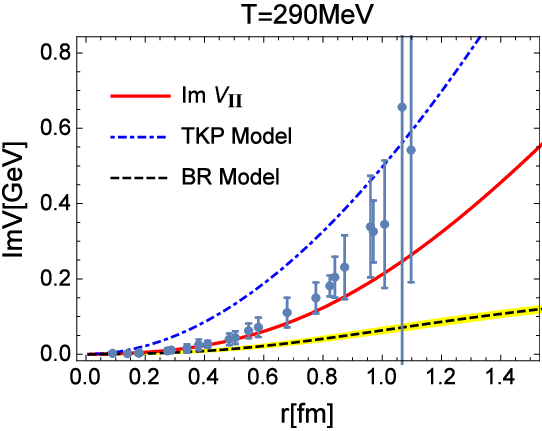}
\includegraphics[width=0.27\textwidth]{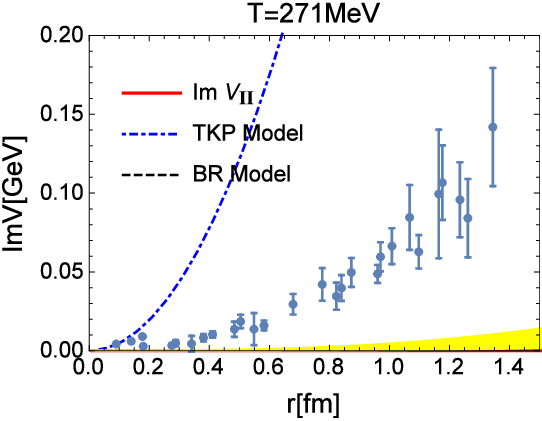}
\vspace*{0.1cm}
\caption{Comparison of $\mathrm{Im}\,V$ between the lattice data in quenched QCD (blue dots) from Ref.~\cite{Burnier:2016mxc} and various complex potential models. The results from the improved KMS potential model $V_{\rm II}$ is denoted by the red solid curve and the blue dash-dotted curve denotes the TKP model. The prediction from BR model is also shown in this figure which is denoted by the black dashed curve with the colored error bands from the uncertainty in the determination of $m_D$. The critical temperature $T_c=290$ ${\rm MeV}$.}\label{compim2}
\end{figure}

The Debye masses at different temperatures as given in Table~\ref{debyemassfit2} are used to evaluate the imaginary part of the HQ potential. The comparisons among various potential models as well as the lattice results are given in Fig.~\ref{compim2}. Below the critical temperature, lattice simulations suggest very small values for ${\rm Im}\,V$ which is in accordance with the numerical evaluations based on the improved KMS model and the BR model. As discussed in Sec.~\ref{imKMS}, due to the elimination of the problematic term $\sim {\hat r}^2$, $\mathrm{Im}\,V_{\rm II}$ actually vanishes at finite $r$ because the Debye mass $m_D\rightarrow 0$.  On the other hand, the TKP model exhibits a rapid increase according to $r^2$ which is the same as the extended KMS model $\mathrm{Im}\,V_{\rm I}$ and qualitatively differs from the lattice data. In the deconfined phase, the predictions from the TKP model overshoot the data in the entire temperature region. Because of the logarithmic divergence at large $r$, the results are even worse as compared to $\mathrm{Im}\,V_{\rm I}$. On the contrary, the BR model underestimates the lattice data at intermediate and large distances and the agreement is only reached for large $T$ and small $r$\cite{Burnier:2016mxc}. On the other hand, for temperatures not far above $T_c$, the results obtained from $\mathrm{Im}\,V_{\rm II}$ point to a significant improvement compared to other two potential models and a quantitative description of the lattice data is achieved. In addition, both the BR model and $\mathrm{Im}\,V_{\rm II}$ asymptotically approach to some constant, however, they present a very different behavior at relatively large distances. The former turns to be saturated much more quickly, while the latter gets a continuous increase even at $r\sim 1$ ${\rm fm}$. In fact, such an increase seems to be consistent with the lattice data although the error bars get large in this distance region. Notice that the asymptotic value at infinitely large $r$ in the BR model turns to be much smaller than that in $\mathrm{Im}\,V_{\rm II}$, therefore, the large $r$ behavior of $\mathrm{Im}\,V$ needs to be further confirmed by the lattice.

As the temperature increases, a visible deviation from the data starts to emerge in $\mathrm{Im}\,V_{\rm II}$ at intermediate distances. We would like to mention that in order to have a ``most confining" potential, the parameter $a$ introduced in Eq.~(\ref{modDR2}) is chosen to be $4$ although in principle it can take some other values smaller than $4$. We have checked that if a relatively smaller value is used, the result for $\mathrm{Im}\,V$ at $T=406$ ${\rm MeV}$ can be further improved. This actually suggests using a $T$-dependent parameter $a$ which of course complicates the proposed model and will not be discussed in more details in the current work. On the other hand, the HQ potential at temperatures close to $T_c$ is most relevant for the studies of quarkonia. Only very heavy bound states, such as the ground state of bottomonium, can survive at very high temperatures whose typical root-mean-square radii are small. As a result, one can expect the improved KMS model is sufficient to describe the interforce between the quark and antiquark and can be put into the Schr\"odinger equation to quantitatively study the properties of the quarkonia.

\section{Summary}\label{sum}

In this paper, we proposed a model for the complex HQ potential which is defined as the Fourier transform of the static gluon propagators in the Keldysh representation. These propagators consist of two parts: the Coulombic term comes from the resummed HTL perturbation theory at leading order while the string contributions are induced by the dimension two gluon condensate.  For the retarded/advanced propagator, the nonperturbative contributions were assumed to be $m_G^2/(p^2+m_D^2)^2$ at static limit. The corresponding symmetric propagator was determined by using the relation given in Eq.~(\ref{Rela}) where the $p_0$-dependence of the retarded/advanced propagator needs to be specified. The resulting potential model $V_{\rm I}$ has a real part which is identical to the well-known KMS potential model while the imaginary part exhibits some unexpected behaviors which were demonstrated by inspecting the asymptotic limit when $\hat{r}\rightarrow 0$.

Improvements on the extended KMS potential model $V_{\rm I}$ have been studied by adding an extra non-perturbative term in the static gluon propagator $D_{R/A}$. Based on the same justification for the introduction of $D_{R/A}^{\rm np}$ in Eq.~(\ref{fullDR}), the new added term $\delta D_R \sim m_G^2 m_D^2/(p^2+m_D^2)^3$ also arises as the consequence of the dimension two gluon condensate. A specific $p_0$-dependence in $\delta D_R$ is adopted for the purpose of eliminating the unwanted $\sim \hat{r}^2$ term appearing in $\mathrm{Im}\,V_{\rm I}^{\rm np}$ in the small $\hat{r}$ limit. The improved KMS potential model $V_{\rm II}$ presents the correct asymptotic behaviors and also reproduces the results from the lattice simulations on the complex HQ potential fairly well.

The comparisons among different potential models show that a satisfactory prediction on the lattice data for $\mathrm{Re}\,V$ seems not very sensitive to the exact forms used in these models. Therefore, modeling $\mathrm{Im}\,V$ turns to be the most challenge task. In the TKP model and BR model, employing a complex dielectric function naturally gives rise to an imaginary part in the HQ potential. Such a dielectric function computed within the HTL perturbation theory at leading order may not capture the full nonperturbative effects due to the low frequency modes. A more consistent way to model the medium effects relies on a nonperturbative evaluation on the dielectric function. On the other hand, the imaginary part in the improved KMS potential model comes from the Fourier transform of the symmetric gluon propagator and the corresponding nonperturbative contributions are determined through a phenomenological gluon propagator which is induced by the dimensional two gluon condensate and has been widely used in some other studies before. We emphasize that in our approach, the $p_0$-dependence of the (nonperturbative) retarded/advanced gluon propagator has to be specified which is crucial to determine the symmetric propagator and then $\mathrm{Im}\,V$. By mimicking the $p_0$-dependence of the perturbative propagator which is known at the fundamental level, a similar $p_0$-dependence was introduced in the nonperturbative counterpart where an extra constraint has also been taken into account in Sec.~\ref{imKMS} in order to ensure a correct asymptotic behavior.

Our complex HQ potential model as given by Eq.~(\ref{imV}) takes a relatively simple form. The real part reproduces the Cornell potential in the confined phase where $m_D\rightarrow 0$, while in the deconfined phase it gets screened for both Coulombic and linear rising string contributions. The imaginary part develops a nonvanishing contribution above $T_c$ which increases rapidly with the quark pair separation for not very high temperatures. There are three parameters appearing in the potential model where the strong coupling $\alpha_s$ and string tension $\sigma$ are assumed to be $T$-independent and can be determined from lattice simulations at zero-temperature. Therefore, there is only one free parameter $m_D$ related to the hot medium effect. We further considered to extract the Debye mass from the model fit to the in-medium $\mathrm{Re}\,V$ from lattice. The obtained values of $m_D$ have been used to evaluate the imaginary part. The outcome suggests that in a $T$-$r$ region relevant to quarkonium physics, the improved KMS potential model $V_{\rm II}$ proposed in this work clearly shows an improved agreement on the lattice data for $\mathrm{Im}\,V$. Therefore, it can offer a quantitative description of the interquark forces which is important for other phenomenological studies on quarkonia.

Finally, we want to point out that the asymptotic behaviors at very large $r$ change dramatically among different potential models which cannot be judged based on the current data from lattice. Therefore, a more accurate lattice reconstruction of the HQ potential, especially for the imaginary part, is urgently needed which requires lattices with finer spacing and larger volume. It is expected to provide more information to constrain the model construction. The different behaviors predicted by different models also need to be further checked when more data from lattice becomes available. Furthermore, the extension of our model to full QCD is also important where we can naively expect that the medium effects including the thermal activations of the light quarks are entirely encoded in the Debye mass as the quenched case. The determination of the QCD Debye mass is crucial for phenomenology applications to heavy-ion collisions. All of these need to be further investigated in the future work.

{\bf Note added:} In a previous version (arXiv:1806.04376v1), we constructed the complex HQ potential based on a completely different idea where the long distance behavior between the quark pair was considered as an effective one-dimensional string interaction. Therefore, the nonperturbative terms in the potential were determined through the one-dimensional Fourier transform of the resummed gluon propagator at static limit. Although formally the real part of the potential obtained in this way is identical to $\mathrm{Re}\,V_{\rm I}$, the corresponding imaginary part asymptotically has a $\sim {\hat r}^2$ contribution similar as $\mathrm{Im}\,V_{\rm I}$ as well as a logarithmic divergence just like the TKP model. The incorrect behavior at small ${\hat r}$ limit has been regulated after a nonperturbative ``entropy" contribution was introduced. However, the logarithmic divergence is still there. When compared with the lattice data, the results from this model are not as good as the one proposed in the current paper. In fact, the idea of an entropy contribution is based on the previous discussions about the real-valued HQ potential. Following the idea in Ref.~\cite{Mocsy:2008eg}, the real-valued potential is considered to lie between the HQ free energy and internal energy, therefore, the entropy contribution should be partially subtracted from the free energy. Of course, this is somehow a vague statement which is not timely anymore due to the already mentioned groundbreaking works\cite{Laine:2006ns,Brambilla:2008cx}. In addition, such a statement loses a clear physical meaning when applied to the imaginary part of the HQ potential. The HQ free energy was originally defined in Ref.~\cite{McLerran:1981pb} which is given by the correlation of two Polyakov loops in the imaginary time formalism. As computed in Ref.~\cite{Petreczky:2005bd}, the singlet free energy at leading order coincides with the real part of the perturbative potential as given in Eq.~(\ref{Vpre}). However, in general, its exact relation to the HQ potential is not clear on the level of an EFT\cite{Brambilla:2010xn,Berwein:2017thy}.

The complex HQ potential model in the current paper was proposed without invoking the above statement about the entropy contribution which puts our model construction on a more solid theoretical footing. In the meantime, it is worthwhile to point out that for our special choose $a=4$, the extra contributions $\mathrm{Re}\,\delta V$ and $\mathrm{Im}\,\delta V$ as discussed in Sec.~\ref{imKMS} are indeed identical to an ``entropy" contribution $-T \partial/\partial T$ where the Debye mass $m_D$ is assumed to linearly depend on $T$ and the derivative acts on the nonperturbative term $\mathrm{Re}\,V_{\rm I}^{{\rm np}}(\hat{r})$ and $\mathrm{Im}\,V_{\rm I}^{{\rm np}}(\hat{r})$, respectively. According to such an interesting finding, it is certainly a meaningful work to explore the relation between the HQ free energy and the potential in the future.

\subsection*{Acknowledgments}
\label{sec:acknowledgments}

We are very grateful to A. Rothkopf for providing us the lattice data of the complex heavy quark potential. We also thank A. Dumitru for helpful comments on the manuscript. This work is supported by the NSFC of China under Project No. 11665008, by Natural Science Foundation of Guangxi Province of China under Project No. 2016GXNSFFA380014, No. 2018GXNSFAA138163 and by the Hundred Talents Plan of Guangxi Province of China. The research of M.M was supported by the European Research Council Grant No. HotLHC ERC-2011-StG-279579.

\end{document}